\documentclass[conference]{IEEEtran}

\usepackage[utf8]{inputenc} % According to the encoding used, you may replace "utf8" with "latin5"
\usepackage[T1]{fontenc}

\usepackage[pdftex]{graphicx}

\usepackage[tight,footnotesize]{subfigure}

\usepackage{cite}

\usepackage[cmex10]{amsmath}
\usepackage{amssymb}
\usepackage{mathrsfs}
\usepackage{siunitx}

\usepackage{url}
\usepackage{lipsum}
\usepackage{multirow}
\usepackage{array}

\newcolumntype{L}[1]{>{\raggedright\let\newline\\\arraybackslash\hspace{0pt}}m{#1}}
\newcolumntype{C}[1]{>{\centering\let\newline\\\arraybackslash\hspace{0pt}}m{#1}}
\newcolumntype{R}[1]{>{\raggedleft\let\newline\\\arraybackslash\hspace{0pt}}m{#1}}

\newcommand{\diCO}{D}

\newcommand{\deltat}{\Delta{t}}
\newcommand{\deltaX}{\Delta{X}}
\newcommand{\deltaY}{\Delta{Y}}
\newcommand{\deltaZ}{\Delta{Z}}

\newcommand{\yakinsama}{\mathscr{N}\text{(}\mu\text{, }\sigma^{\text{2}}\text{)}}

\newcommand{\rarrow}{\overrightarrow{r}}

\usepackage{color}
\usepackage{todonotes}
\definecolor{morange}{rgb}{0.8,0.2,0}
\definecolor{mblue}{rgb}{0,0.3,1.0}
\definecolor{mbluee}{rgb}{0.4,0.1,0.9}
\begin{document}

\title{Channel Model of Molecular Communication via Diffusion in a Vessel-like Environment Considering a Partially Covering Receiver}

% affiliations
\author{\IEEEauthorblockN{Meriç Turan$^{1}$, Mehmet Şükrü Kuran$^{2}$, H. Birkan Yilmaz$^{3}$, Ilker Demirkol$^{3}$, Tuna Tugcu$^{1}$}
\IEEEauthorblockA{$^{1}$Department of Computer Engineering, NETLAB, Bogazici University, Istanbul, Turkey\\
$^{2}$Department of Computer Engineering, Abdullah Gul University, Kayseri, Turkey\\
$^{3}$Department of Telematics Engineering, Universitat Politecnica de Catalunya, Barcelona, Spain\\
{E-mails:\{meric.turan, tugcu\}}@boun.edu.tr, sukru.kuran@agu.edu.tr, birkan.yilmaz@upc.edu, ilker.demirkol@entel.upc.edu
}
}

% conference papers do not typically use \thanks and this command
% is locked out in conference mode. If really needed, such as for
% the acknowledgment of grants, issue a \IEEEoverridecommandlockouts
% after \documentclass

% make the title area
\maketitle

% The abstract should be in passive voice
\begin{abstract}
%\lipsum[1]
By considering potential health problems that a fully covering receiver may cause in vessel-like environments, the implementation of a partially covering receiver is needed. To this end, distribution of hitting location of messenger molecules (MM) is analyzed within the context of molecular communication via diffusion with the aim of channel modeling. The distribution of these MMs for a fully covering receiver is analyzed in two parts: angular and radial dimensions. For the angular distribution analysis, the receiver is divided into 180 slices to analyze the mean, standard deviation, and coefficient of variation of these slices. For the axial distance distribution analysis, Kolmogorov-Smirnov test is applied for different significance levels. Also, two different implementations of the reflection from the vessel surface (i.e., rollback and elastic reflection) are compared and mathematical representation of elastic reflection is given. The results show that MMs have tendency to spread uniformly beyond a certain ratio of the distance to the vessel radius. By utilizing the uniformity, we propose a channel model for the partially covering receiver in vessel-like environments and validate the proposed model by simulations.

\end{abstract}


\IEEEpeerreviewmaketitle

%%%%%%%%%%%%%%%%%%%%%%%%%%%%%%%%%%%%%%%%%%%%%%%%%%%%
%%%%%%%%%%%%%%%%%%%%%%%%%%%%%%%%%%%%%%%%%%%%%%%%%%%%
%%%%%%%%%%%%%%%%%%%%%%%%%%%%%%%%%%%%%%%%%%%%%%%%%%%%
%%%%%%%%%%%%%%%%%%%%%%%%%%%%%%%%%%%%%%%%%%%%%%%%%%%%
\section{Introduction}

% - General introduction to MCvD
Molecular communication via diffusion (MCvD) is one of the  promising molecular communication (MC) systems proposed in the context of nanonetworking, especially for \textit{in vivo} applications. In contrast to the classical communication systems, MCvD utilizes molecules, messenger molecules (MMs) as information carriers, mainly for high bio-compatibility and energy efficiency. These molecules propagate in a fluid environment (e.g., inter-cellular fluid) and follow Brownian motion, which has vastly different features than the physical layers of classical communication systems~\cite{Farsad2016_Comprehensive}.

% - Free-diffusion environment VS vessel-like environment
Most of the research conducted regarding MCvD assumes a free diffusion environment in which the MMs can move in an unconstrained manner. This free diffusion environment, while being very suitable for developing mathematical channel, noise, and interference models, has several shortcomings such as limited range and limited applicability to \textit{in vivo} environments. As shown in previous works \cite{Kuran2010_Energy, Yilmaz2014_Simulation, Yilmaz2014_ThreeDim, Farsad2016_Comprehensive}, the performance of the MCvD system sharply decreases as the distance between the transmitting pair exceeds several tens of micrometers. This limited range severely hinders the applicability of MCvD in a free diffusion environment without any enhancements. Also, \textit{in vivo} environments inside complex living organisms mainly consist of huge bodies of cells next to each other or include vessel-like environments, which are different than the unbounded and unconstrained free diffusion environments used in the MCvD literature.

% - Features of the vessel-like environment as well as the receiver design
As an alternative to the free diffusion environment, a tunnel-like environment which closely resembles the inside of a blood vessel, has been proposed in several works in the literature \cite{Farsad2012_Onchip,Kuran2013_Tunnel,Azadi2016_Novel, Wicke2017_Modeling, TURAN2018_MOLEye}. Unlike the free diffusion environment, in the vessel-like environment the MMs are bounded by the walls of the vessel, and upon impact to these walls they are either reflected back or absorbed, depending on the environmental model. Since these walls constraint the movement of the MMs, the effective range of the system greatly increases.

In this work, we study the channel model of such a vessel-like environment considering a receiver that partially covers the cross-section of the vessel and vessel walls that reflect the MMs upon contact. In our analysis, we choose a partially covering receiver over a fully covering receiver for increased bio-compatibility. We assume devices that are just utilizing already existing tunnels for an artificial purpose (e.g., detecting lipids, bio-markers, or blood clots inside blood-vessels).
%%%%%%%%%%%%%%%%%%%%%%%%%%%%%%%%%%%%%%%%%
\begin{figure*}[!t]
	\begin{center}
    \subfigure[Rollback strategy]
        {\includegraphics[width=0.98\columnwidth,keepaspectratio]%
		{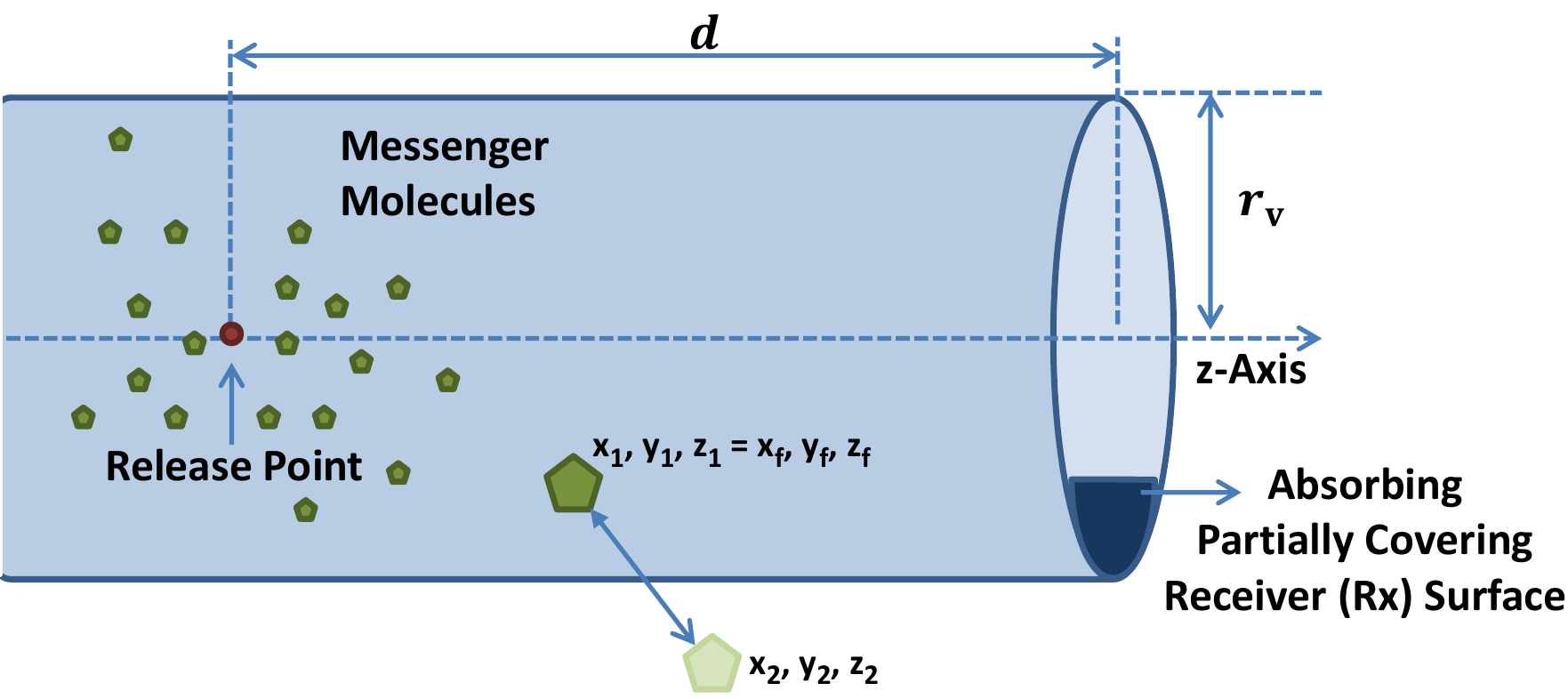}
        \label{fig_rollbacking_communication_model}}    \subfigure[Elastic reflection strategy]
		{\includegraphics[width=0.98\columnwidth,keepaspectratio]%
		{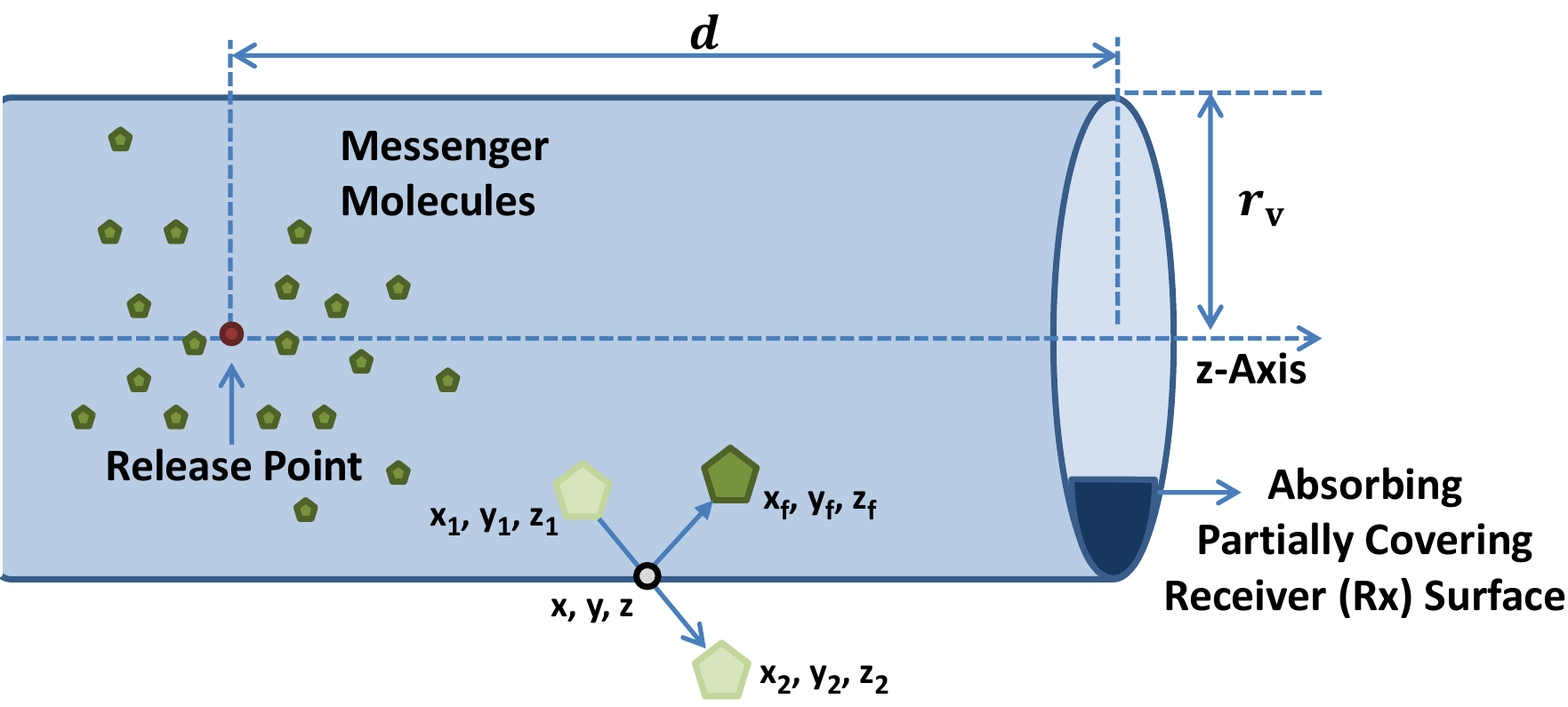}
		\label{fig_reflective_communication_model}}

	\caption{Micro-fluidic communication channel model and reflection strategies for partially covering and fully reflective boundary surfaces}
	\label{fig_communication_model}
	\end{center}
\end{figure*}
%%%%%%%%%%%%%%%%%%%%%%%%%%%%%%%%%%%%%%%%%%%%%%%%%%%%

% - Contribution of the paper, itemized
The main contributions of the paper are summarized below:
\begin{itemize}
\item We describe and elaborate on two reflection strategies that can be used to simulate Brownian motion in a vessel-like environment, namely: elastic reflection strategy and rollback strategy.
\item Using simulations, we show that the molecular hitting location distribution of MCvD to the cross-section of the vessel follows uniform distributions in both angular and distance-from-the-central-axis dimensions independently beyond a certain ratio of the distance to the vessel radius.
\item Based on the previous observation, we evaluate the hitting rate of molecules to a partially covering receiver as a function of the area of the receiver, the transmitter-receiver distance, and the radius of the vessel.
\end{itemize}

%%%%%%%%%%%%%%%%%%%%%%%%%%%%%%%%%%%%%%%%%%%%%%%%%%%%
%%%%%%%%%%%%%%%%%%%%%%%%%%%%%%%%%%%%%%%%%%%%%%%%%%%%
%%%%%%%%%%%%%%%%%%%%%%%%%%%%%%%%%%%%%%%%%%%%%%%%%%%%
%%%%%%%%%%%%%%%%%%%%%%%%%%%%%%%%%%%%%%%%%%%%%%%%%%%%
\section{System Model}
Most of the prior work in the MC literature consider free diffusion environment where the MMs can roam freely without any boundaries in the environment. In contrast, we consider cylindrical (i.e., vessel-like) environment in this work. This vessel-like environment is more suitable to model significant \textit{in vivo} and \textit{in vitro} applications, e.g., sensing applications in blood vessels of a human body and micro-fluidic channels.

%%%%%%%%%%%%%%%%%%%%%%%%%%%%%%%%%%%%%%%%%%%%%%%%%%%%
\subsection{Diffusion Model}
We consider a diffusion model consisting of a point transmitter, a fully absorbing circular receiver, a single type of information carrying MM, and a vessel-like environment. The vessel-like environment is considered to be a perfect cylinder with a fully reflecting surface as in Fig.~\ref{fig_communication_model}. 

In the diffusion model, the total displacement along the x-axis ($\deltaX$) of an MM in $\deltat$ duration follows a Gaussian distribution as
%%%%%%%%%%%%%%%%%%%%%%%%%%%%%%%%%%%%%%%%%%%%%%%%%%%%
\begin{align}
\deltaX \sim \mathscr{N}(0, 2D\Delta t)
\label{eq_displacementX_diffusion}
\end{align}
%%%%%%%%%%%%%%%%%%%%%%%%%%%%%%%%%%%%%%%%%%%%%%%%%%%%
where $\Delta t$ is the simulation time step, $\diCO$ is the diffusion coefficient,  and $\yakinsama$ is the Gaussian random variable with mean $\mu$ and variance $\sigma^2$. Considering the movement in all three axes, the total displacement in a single time step is calculated as
%%%%%%%%%%%%%%%%%%%%%%%%%%%%%%%%%%%%%%%%%%%%%%%%%%%%
\begin{align}
\rarrow = (\deltaX, \deltaY, \deltaZ)
\label{eq_displacementXYZ}
\end{align}
%%%%%%%%%%%%%%%%%%%%%%%%%%%%%%%%%%%%%%%%%%%%%%%%%%%%
where $\deltaY$ and $\deltaZ$ correspond to the displacement in the y- and the z-axes, respectively, both of which follow a Gaussian distribution with the same $\mu$ and $\sigma^2$ values in~\eqref{eq_displacementX_diffusion}.

%%%%%%%%%%%%%%%%%%%%%%%%%%%%%%%%%%%%%%%%%%%%%%%%%%%%
%%%%%%%%%%%%%%%%%%%%%%%%%%%%%%%%%%%%%%%%%%%%%%%%%%%%
\subsection{Simulating Diffusion near Reflective Vessel Surface}
There are two common simulation implementations for the fully reflective vessel surface in the MC literature, namely rollback and elastic reflection strategies. The most commonly used one is the rollback strategy. In this approach, the molecules that hit a surface roll back as the name suggests (Fig.~\ref{fig_rollbacking_communication_model}). It is easier to implement and faster run times can be achieved when the reflection strategy is chosen as rollback. However, there is a trade off between complexity and accuracy. It is less realistic to implement a cell or vessel reflection strategy as rollback.

The second approach is to implement reflection strategy as elastic reflection. Molecules that hit a reflective surface make perfectly elastic collision in elastic reflection strategy. Due to its accuracy, we utilize elastic reflection in our simulations. In the rest of this section, mathematical representation of the elastic reflection implementation is described.

In our topology, there are two intersection points between the path of a molecule (represented as a line) and a cylinder. Since net displacement in the z-axis is the same after reflection, we can ignore the movement in the z-axis. Therefore, the equation of the cylinder can be written as
%%%%%%%%%%%%%%%%%%%%%%%%%%%%%%%%%%%%%%%%%%%%%%%%%%%%
\begin{align}
(x - x_3)^2 + (y - y_3)^2 = r_{\text{v}}^2
\label{eq_cylinderEquation}
\end{align}
%%%%%%%%%%%%%%%%%%%%%%%%%%%%%%%%%%%%%%%%%%%%%%%%%%%%
where $r_{\text{v}}$ is the radius of the vessel (cylinder), $(x, y)$ is the intersection point, and $(x_3, y_3)$ is the center of the cylinder. The line that passes through the intersection points follows the line equation
%%%%%%%%%%%%%%%%%%%%%%%%%%%%%%%%%%%%%%%%%%%%%%%%%%%%
\begin{align}
\begin{split}
x &= x_1 + (x_2 - x_1)t\\
y &= y_1 + (y_2 - y_1)t
%        &= \vflow \, \deltat + \deltaXsub{diffusion}
\end{split}
\label{eq_lineEquation}
\end{align}
%%%%%%%%%%%%%%%%%%%%%%%%%%%%%%%%%%%%%%%%%%%%%%%%%%%%
where $(x_1, y_1)$ is the location of the molecules at time $m\Delta t$, and $(x_2, y_2)$ is the location of the molecules at time $(m+1)\deltat$ if vessel had no boundaries. When the x and y equations are placed in the cylinder equation, we obtain
%%%%%%%%%%%%%%%%%%%%%%%%%%%%%%%%%%%%%%%%%%%%%%%%%%%%
\begin{align}
at^2 + bt + c = 0
\label{eq_quadraticEquation}
\end{align}
%%%%%%%%%%%%%%%%%%%%%%%%%%%%%%%%%%%%%%%%%%%%%%%%%%%%
where a, b, and c are
%%%%%%%%%%%%%%%%%%%%%%%%%%%%%%%%%%%%%%%%%%%%%%%%%%%%
\begin{align}
\begin{split}
a &= (x_2 - x_1)^2 + (y_2 - y_1)^2\\
b &= 2[(x_2 - x_1)(x_1 - x_3) + (y_2 - y_1)(y_1 - y_3)]\\
c &= {x_3}^2 + {y_3}^2 + {x_1}^2 + {y_1}^2 - (x_3x_1 + y_3y_1) - r_{\text{v}}^2 \, .
\end{split}
\label{eq_abcValue}
\end{align}
%%%%%%%%%%%%%%%%%%%%%%%%%%%%%%%%%%%%%%%%%%%%%%%%%%%%
Then, the solution for t can be evaluated by finding the roots of \eqref{eq_quadraticEquation}. The intersection points can be calculated by substituting $t$ in (\ref{eq_lineEquation}) by finding the roots of $t$. Since there are two intersection points, the one closer to the last position of the molecule should be selected as the actual intersection point. After finding the intersection point, the position of the molecule after collision can be found as
%%%%%%%%%%%%%%%%%%%%%%%%%%%%%%%%%%%%%%%%%%%%%%%%%%%%
\begin{align}
\begin{split}
x_f &= 2x - x_2\\
y_f &= 2y - y_2\\
z_f &= z_2
\end{split}
\label{eq_moleculeNewPosition}
\end{align}
%%%%%%%%%%%%%%%%%%%%%%%%%%%%%%%%%%%%%%%%%%%%%%%%%%%%
where $(x_f, y_f, z_f)$ is the location of the molecule after reflection occurred, and $z_2$ is the location of the molecule's z-axis at time $(m+1)\Delta t$ if vessel had no boundaries.

All in all, perfectly elastic collision is realized in this work due to its rationality despite the complexity and longer simulation time. Note that this work only considers single collision. In other words, $\deltat$ should be small enough and $r_{\text{v}}$ should be large enough to avoid multiple bounces in a single step. Yet, simulation using excessively large $\deltat$ causes unreliable results even for the no boundaries case. Also, putting nanomachines into extra slim vessels (at least 10 times thinner than capillaries in terms of radius~\cite{Hall2015_Medical}) may cause vascular occlusion. Therefore, accuracy of the simulation has much troubled problems in such case.

%All in all, perfectly elastic collision is realized in this work due to its rationality despite the complexity and longer simulation time. Note that this work only considers single collision. In other words, if $\deltat$ is too large or $r_{\text{v}}$ is too small, which may cause multiple bouncing in a single $\Delta t$, then this formula does not work. Yet, simulation using excessive $\Delta t$ causes unreliable results even for the no boundaries case. Also, putting nanomachines into extra slim vessels (at least 10 times thinner than capillaries in terms of radius~\cite{Hall2015_Medical}) may cause vascular occlusion. Therefore, accuracy of the simulation has much troubled problems in such case.

%%%%%%%%%%%%%%%%%%%%%%%%%%%%%%%%%%%%%%%%%%%%%%%%%%%%%%%
%%%%%%%%%%%%%%%%%%%%%%%%%%%%%%%%%%%%%%%%%%%%%%%%%%%%%%%
%%%%%%%%%%%%%%%%%%%%%%%%%%%%%%%%%%%%%%%%%%%%%%%%%%%%%%%
%%%%%%%%%%%%%%%%%%%%%%%%%%%%%%%%%%%%%%%%%%%%%%%%%%%%%%%
\section{Distribution Analysis of Hitting Location}
The results presented in this section are obtained from the custom-made simulator that keeps track of the location of the molecules and stores the molecules that arrive at the destination. By utilizing the simulation output, we evaluate the overall distribution of the location of the received molecules under different environmental conditions with the goal of simplifying the channel model via homogeneity of the molecule hitting locations. We consider MCvD in a vessel-like environment as depicted in Fig.~\ref{fig_reflective_communication_model} with fully covering receiver and the system parameters are given in Table~\ref{simulation_parameters}.
%%%%%%%%%%%%%%%%%%%%%%%%%%%%%%%%%%%%%%%%%%%%%%%%%%%%
\begin{table}[th]
% increase table row spacing, adjust to taste
\renewcommand{\arraystretch}{1.2}
\caption{Simulation parameters}
\label{simulation_parameters}
\begin{center}
\begin{tabular}{ L{5cm}  L{3cm} }
  \hline			
  Parameter & Value \\
  \hline			  
  Radius of vessel ($r_{\text{v}}$) & $2\sim 5 \si{\micro\metre}$ \\
  Distance between Tx and Rx ($d$) & $3.4\sim 10 \si{\micro\metre}$ \\
  Diffusion coefficient ($\diCO$) & $\{100, 200, 400 \}\si{\micro\metre^{2}\per\second}$ \\
  Simulation time step ($\deltat$)    & \SI{0.1}{\milli\second} \\
  Number of released molecules ($N^{\text{Tx}}$)       & \SI{1.5} million \\ 
  \hline  
\end{tabular}
\end{center}
\end{table}
%%%%%%%%%%%%%%%%%%%%%%%%%%%%%%%%%%%%%%%%%%%%%%%%%%%%

When the circular receiver fully covers the vessel in two-dimensions, an MM hits the receiver and the hitting point has a certain distance and an angle according to the center of the receiver. In order to show the tendency of the molecules to be distributed uniformly in the 3D environment, we analyze the distribution in terms of both the angle and the distance of the molecules with respect to the center of the receiver. There are three dimensions while analyzing the hitting location distribution:
\begin{itemize}
\item \textit{Distribution in x and y axis}: These two distributions can be packed into an angular distribution.
\item \textit{Distribution in the radius}: After the distribution in the angle is analyzed, the remaining parameter is the distance between the hitting molecules and the center of the receiver (i.e., axial distance). The distribution of axial distances should be uniform between $0$ and $r_{\text{v}}$.
\end{itemize}

By considering these three distributions in two parts, we consider the overall distribution.
%%%%%%%%%%%%%%%%%%%%%%%%%%%%%%%%%%%%%%%%%%%%%%%%%%%%
\begin{figure*}[!t]
	\begin{center}
    \subfigure[Good Environment]
		{\includegraphics[width=0.66\columnwidth,keepaspectratio]%
		{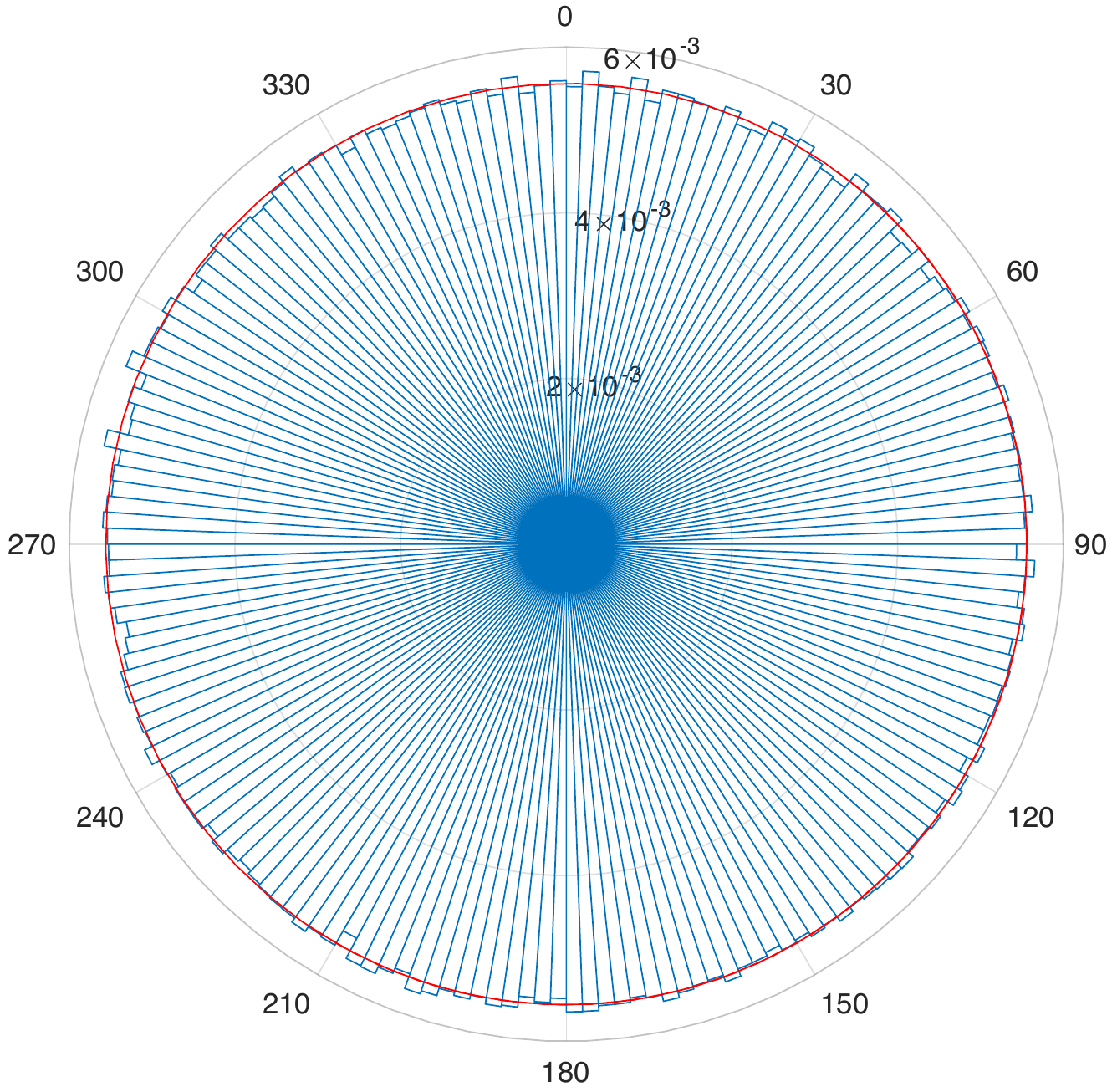}
		\label{fig_hittingMap_D400_d9_r3}}
    \subfigure[Moderate environment]
        {\includegraphics[width=0.66\columnwidth,keepaspectratio]%
		{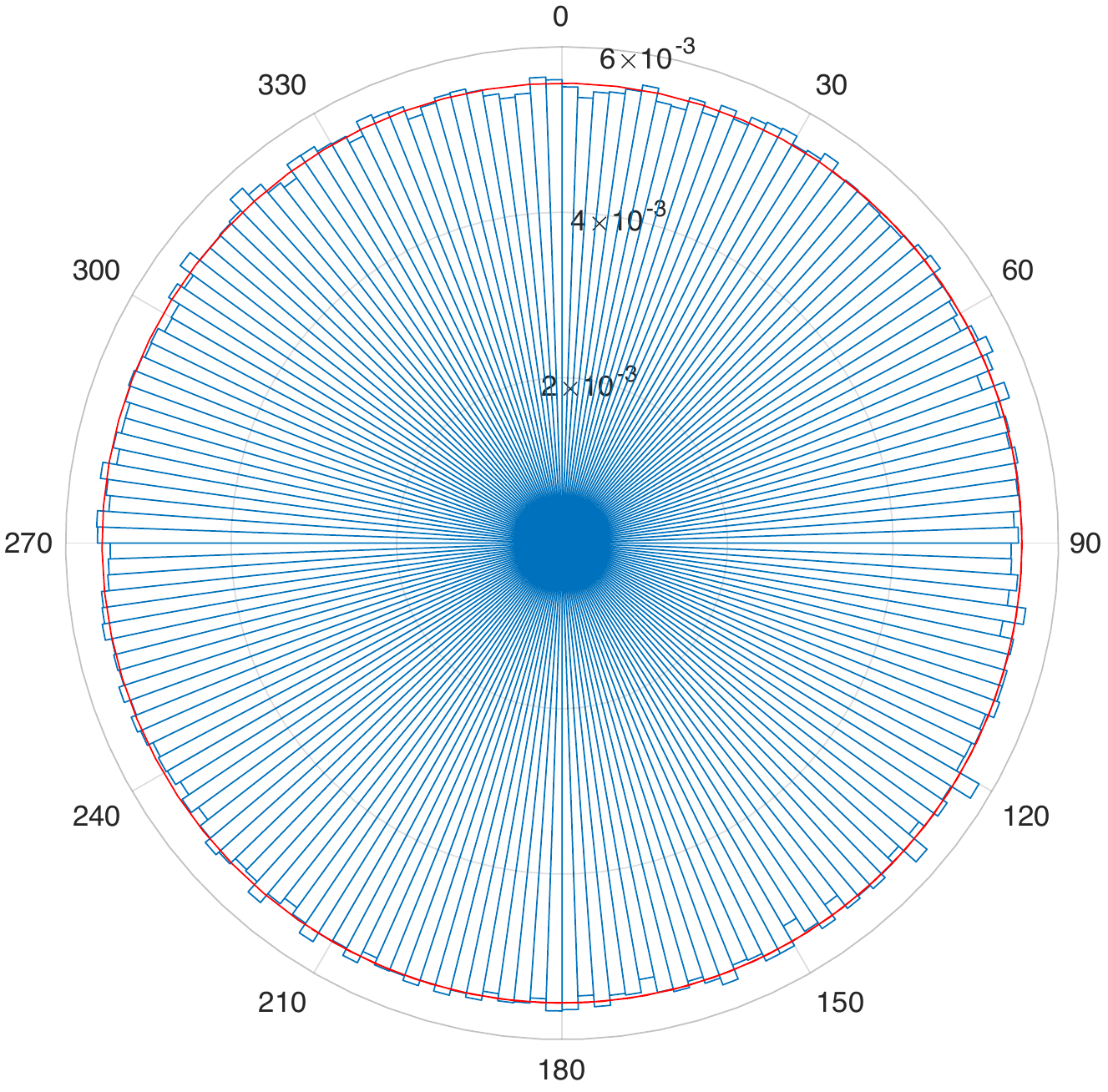}
        \label{fig_hittingMap_D200_d8_r4}}
    \subfigure[Harsh environment]
        {\includegraphics[width=0.66\columnwidth,keepaspectratio]%
		{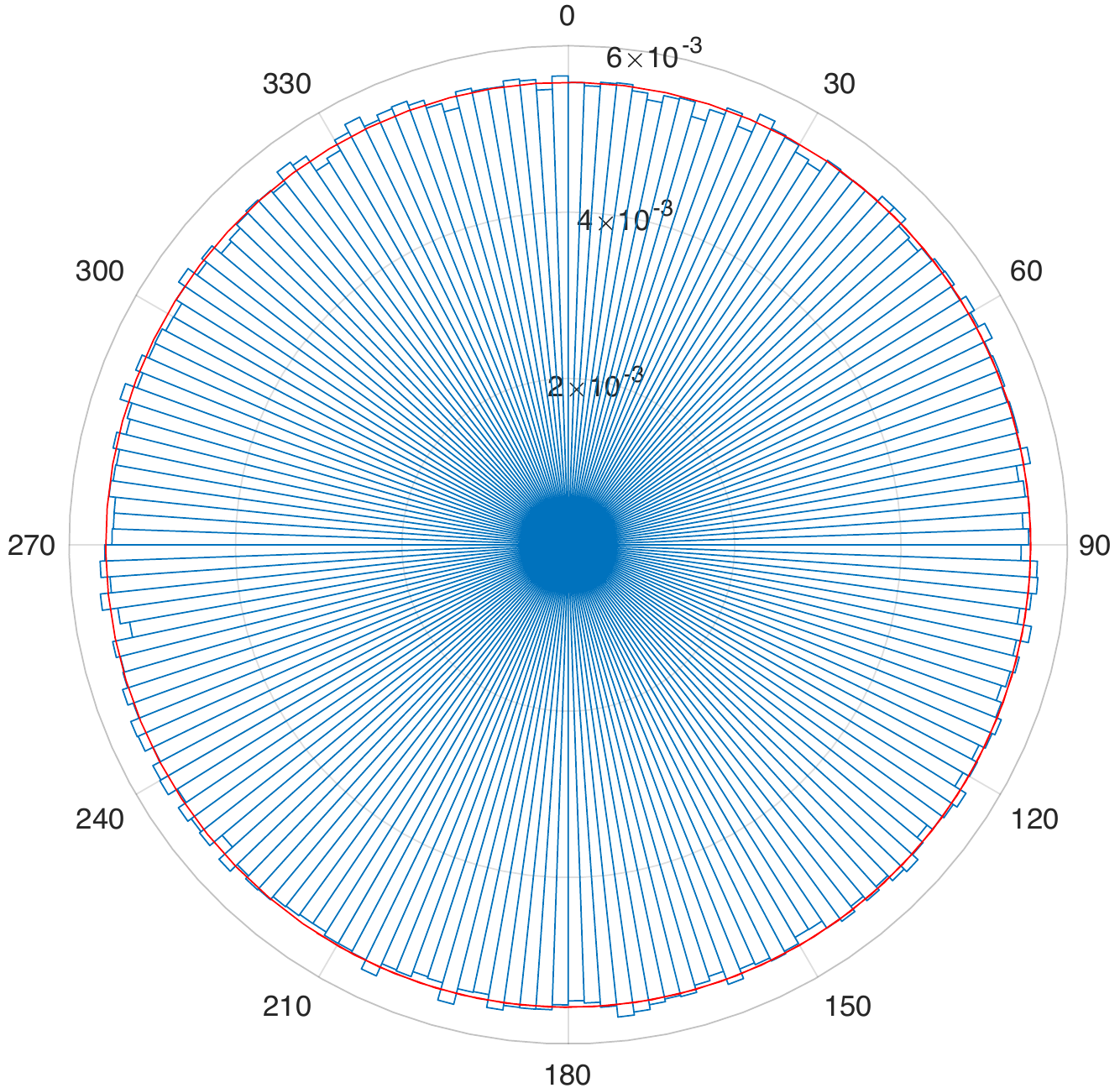}
        \label{fig_hittingMap_D100_d9_r3}}        
	\end{center}
   \caption{Angular distributions of the received molecules under different environmental conditions. Blue lines represents the density of the total number of hitting molecules in that particular slice whereas red lines represent the mean densities. Note that the numbers outside and within the circle represent the degrees and the density of the slices, respectively. Also, for the sake of uniformity in angle, note that the coefficient of variations (Table~\ref{tbl_hittingMap}) are very small.}
   \label{fig_angular_distribution}
\end{figure*}
%%%%%%%%%%%%%%%%%%%%%%%%%%%%%%%%%%%%%%%%%%%%%%%%%%%%

%%%%%%%%%%%%%%%%%%%%%%%%%%%%%%%%%%%%%%%%%%%%%%%%%%%%
%%%%%%%%%%%%%%%%%%%%%%%%%%%%%%%%%%%%%%%%%%%%%%%%%%%%
\subsection{Uniformity in Angle}
Since the receiver is expected to be a circle in the vessel-like environments, angular distribution becomes an important factor while analyzing the distribution of the received molecules. While analyzing uniformity in angle, parameters in Table~\ref{simulation_parameters} are used. In order to analyze the distribution of the received molecules in angle, we additionally define three different environmental conditions by altering $D$ inspired by~\cite{Kuran2010_Energy}, $d$ and $r_{\text{v}}$ inspired by the thinnest part of the capillaries~\cite{Hall2015_Medical}. The environments defined as good, moderate, and harsh by different parameter values are presented in Table~\ref{tbl_env_params}.

%%%%%%%%%%%%%%%%%%%%%%%%%%%%%%%%%%%%%%%%%%%%%%%%%%%%
\begin{table}[th]
\renewcommand{\arraystretch}{1.2}
\caption{Environment Parameters for Distribution Analysis}
\label{tbl_env_params}
\centering
%\begin{center}
  \begin{tabular}{C{2cm} C{1.3cm} C{1.3cm} C{1.3cm}}
    \hline
    Environments & $D (\si{\micro\meter^2/\second})$ & $d (\si{\micro\meter})$ & $r_{\text{v}} (\si{\micro\meter})$ \\ \hline
    Good  & 400 & 7 & 5 \\
    Moderate & 200 & 8 & 4 \\
    Harsh & 100 & 9 & 3 \\
    \hline
  \end{tabular}
%\end{center}
\end{table}
%%%%%%%%%%%%%%%%%%%%%%%%%%%%%%%%%%%%%%%%%%%%%%%%%%%%

Since the number of received molecules are very high, circular receiver is sliced into 180 parts as in Fig.~\ref{fig_angular_distribution}, and the hitting frequencies of these slices are analyzed. Since the molecules move similarly in each direction, the concentration of the slices are expected to be uniform. In that sense, mean, standard deviation, and coefficient of variation of the slices for three different environmental conditions (good, moderate, and harsh) are calculated and given in Table~\ref{tbl_hittingMap}. As seen, the coefficient of variation values are excessively low, which means that the deviations from the mean value are negligible, i.e., the molecules have tendency to spread uniformly with respect to angle. Also, it is shown that the angular distribution is independent from the environmental conditions.
%%%%%%%%%%%%%%%%%%%%%%%%%%%%%%%%%%%%%%%%%%%%%%%%%%%%
\begin{table}[th]
\renewcommand{\arraystretch}{1.2}
\caption{Metrics of Angular Distribution (180 slices)}
\label{tbl_hittingMap}
\centering
%\begin{center}
  \begin{tabular}{C{2.14cm} C{1.7cm} C{1.7cm} C{1.7cm}}
    \hline
    Time & Good Environment & Moderate Environment & Harsh Environment \\ \hline
    Mean  & $5.56  \times 10^{-3}$ & $5.56 \times 10^{-3}$ & $5.56 \times 10^{-3}$ \\
    Standard deviation & $6.44\times10^{-5}$ & $7.72\times10^{-5}$ & $6.99\times10^{-5}$ \\
    Coeff. of variation & $1.16\times10^{-2}$ & $1.39\times10^{-2}$ & $1.26\times10^{-2}$ \\
%    Maximum density & $2.96x10^{-3}$ & $2.94x10^{-3}$ & $2.92x10^{-3}$ \\
%    Minimum density & $2.65x10^{-3}$ & $2.62x10^{-3}$ & $2.63x10^{-3}$ \\
%    Ratio & 10.06\% & 12.18\% & 10.70\% \\
    \hline
  \end{tabular}
%\end{center}
\end{table}
%%%%%%%%%%%%%%%%%%%%%%%%%%%%%%%%%%%%%%%%%%%%%%%%%%%%

%%%%%%%%%%%%%%%%%%%%%%%%%%%%%%%%%%%%%%%%%%%%%%%%%%%%
%%%%%%%%%%%%%%%%%%%%%%%%%%%%%%%%%%%%%%%%%%%%%%%%%%%%
%%%%%%%%%%%%%%%%%%%%%%%%%%%%%%%%%%%%%%%%%%%%%%%%%%%%
\subsection{Uniformity in Radius}
Even if the angular distribution is uniform, we also need to consider the uniformity in the axial distance dimension. There may be a case where the slices in angular dimension have uniform structure but the hitting molecules are mostly close to the center. Therefore, considering all the dimensions is crucial for our modeling purposes.

The ratio of the hitting molecules whose axial distances are less than an arbitrary radius $r_a$ to all hitting molecules within a limited time is denoted by $\mathtt{HR}(r_a|t)$, and 
%%%%%%%%%%%%%%%%%%%%%%%%%%%%%%%%%%%%%%%%%%%%%%%%%%%%
\begin{align}
\mathtt{HR(r_a|t)} = N^{\text{Rx}}(r_a|t) / N^{\text{Rx}}(r_{\text{v}}|t)
\label{eq_hr}
\end{align}
%%%%%%%%%%%%%%%%%%%%%%%%%%%%%%%%%%%%%%%%%%%%%%%%%%%%
where $N^{\text{Rx}}(r|t)$ is the total number of hitting molecules whose axial distances are less than $r$. Note that for the perfect uniformity, we expect that the molecules hit at each point on the surface with equal probability, i.e., $\mathtt{HR(r_a|t)} = (r_a / r_{\text{v}})^2$.

In order to find out whether the received molecules are spread uniformly or not, one sample Kolmogorov-Smirnov (K-S) test is applied by comparing the perfect uniformity case with the empirical distribution function from simulations. For the significance levels of the K-S test, 5\% and 1\% are used. Moreover, in order to make a fair comparison between environments with different diffusion coefficients, multiples of peak times ($t_p$) are used as the time parameter where $t_p = \frac{d^2}{6D}$~\cite{Schulten2000_Lectures}.

As the result of the simulations under three different diffusion coefficients, five different radii of the vessel, and 15 different distances, we find that all the test scenarios pass the K-S test when $d/r_{\text{v}}$ is greater than a specific value (Table~\ref{tbl_ks_tests}). 
%%%%%%%%%%%%%%%%%%%%%%%%%%%%%%%%%%%%%%%%%%%%%%%%%%%%
\begin{table}[t]
\renewcommand{\arraystretch}{1.3}
\caption{K-S Test Results}
\label{tbl_ks_tests}
\centering
%\begin{center}
  \begin{tabular}{C{2.45cm} C{2.45cm} C{2.45cm}}
    \hline
    Symbol duration & 1\% significance level & 5\% significance level \\ \hline
    1-peak time  & $d/r_{\text{v}}\geq$2.00 & $d/r_{\text{v}}\geq$2.40 \\
    2-peak time & $d/r_{\text{v}}\geq$1.92 & $d/r_{\text{v}}\geq$2.00 \\
    3-peak time & $d/r_{\text{v}}\geq$1.88 & $d/r_{\text{v}}\geq$1.98 \\
    5-peak time and over & $d/r_{\text{v}}\geq$1.82 & $d/r_{\text{v}}\geq$1.90 \\
    \hline
  \end{tabular}
%\end{center}
\end{table}

%%%%%%%%%%%%%%%%%%%%%%%%%%%%%%%%%%%%%%%%%%%%%%%%%%%%
%%%%%%%%%%%%%%%%%%%%%%%%%%%%%%%%%%%%%%%%%%%%%%%%%%%%
%%%%%%%%%%%%%%%%%%%%%%%%%%%%%%%%%%%%%%%%%%%%%%%%%%%%
%%%%%%%%%%%%%%%%%%%%%%%%%%%%%%%%%%%%%%%%%%%%%%%%%%%%
\section{Partially Covering Receiver}
%%%%%%%%%%%%%%%%%%%%%%%%%%%%%%%%%%%%%%%%%%%%%%%%%%%%
\begin{figure*}[!t]
	\begin{center}
    \subfigure[$D=200$, $r_{v}=3$, $d=9$]
		{\includegraphics[width=0.66\columnwidth,keepaspectratio]%
		{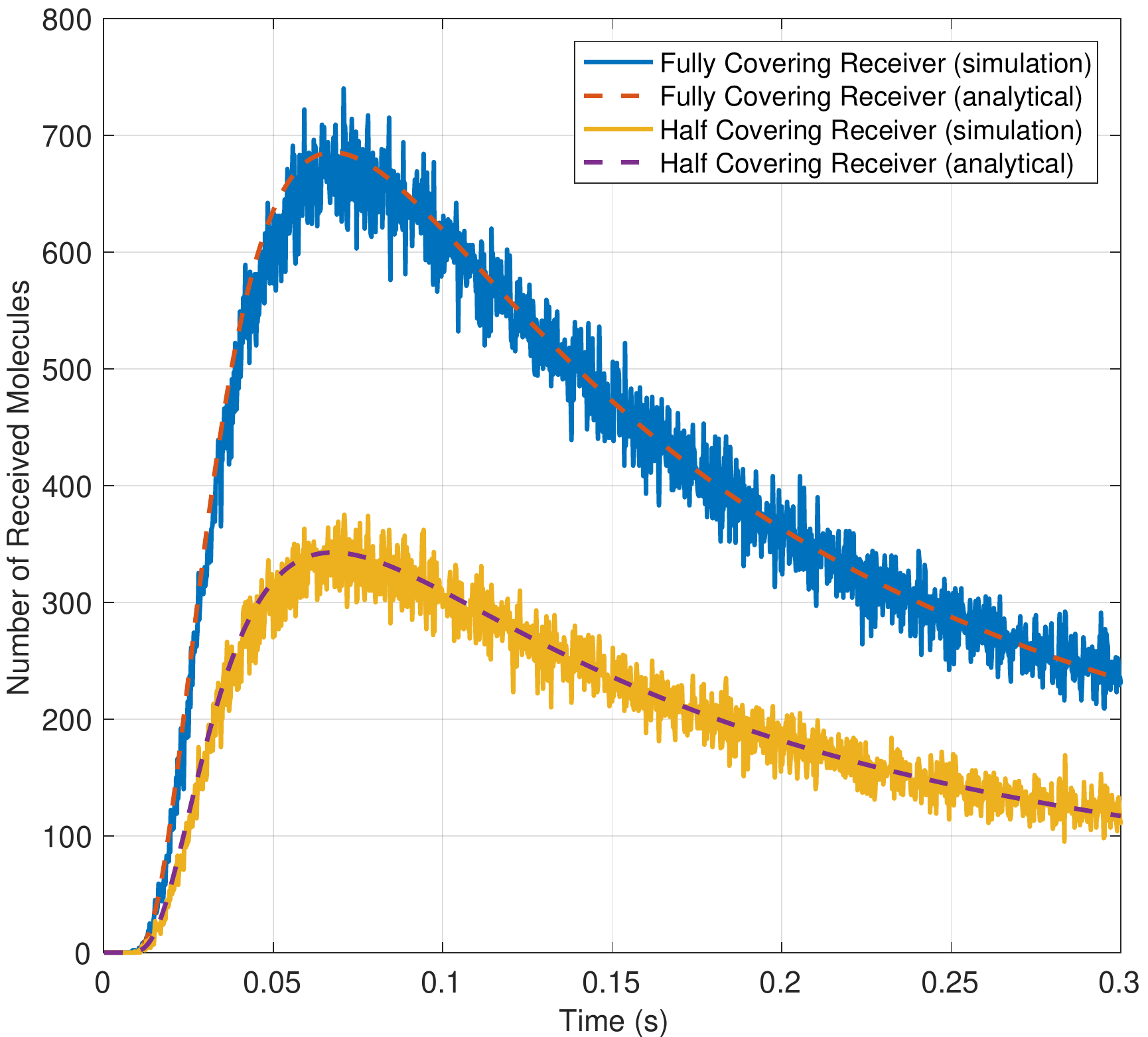}
		\label{fig_patch_r3_d9_D200}}
    \subfigure[$D=200$, $r_{v}=3$, $d=6$]
        {\includegraphics[width=0.66\columnwidth,keepaspectratio]%
		{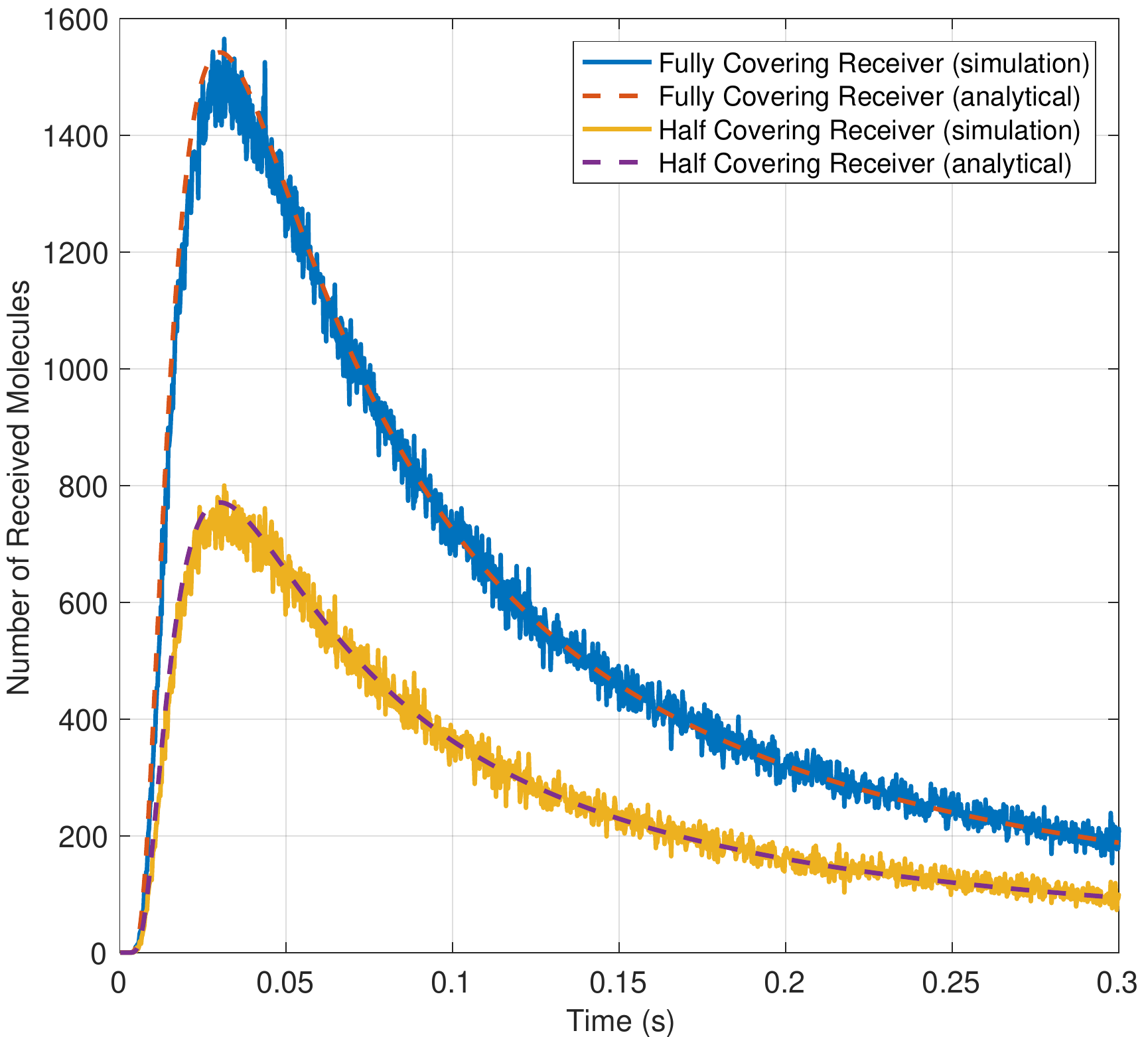}
        \label{fig_patch_r3_d6_D200}}
    \subfigure[$D=100$, $r_{v}=3$, $d=6$]
        {\includegraphics[width=0.66\columnwidth,keepaspectratio]%
		{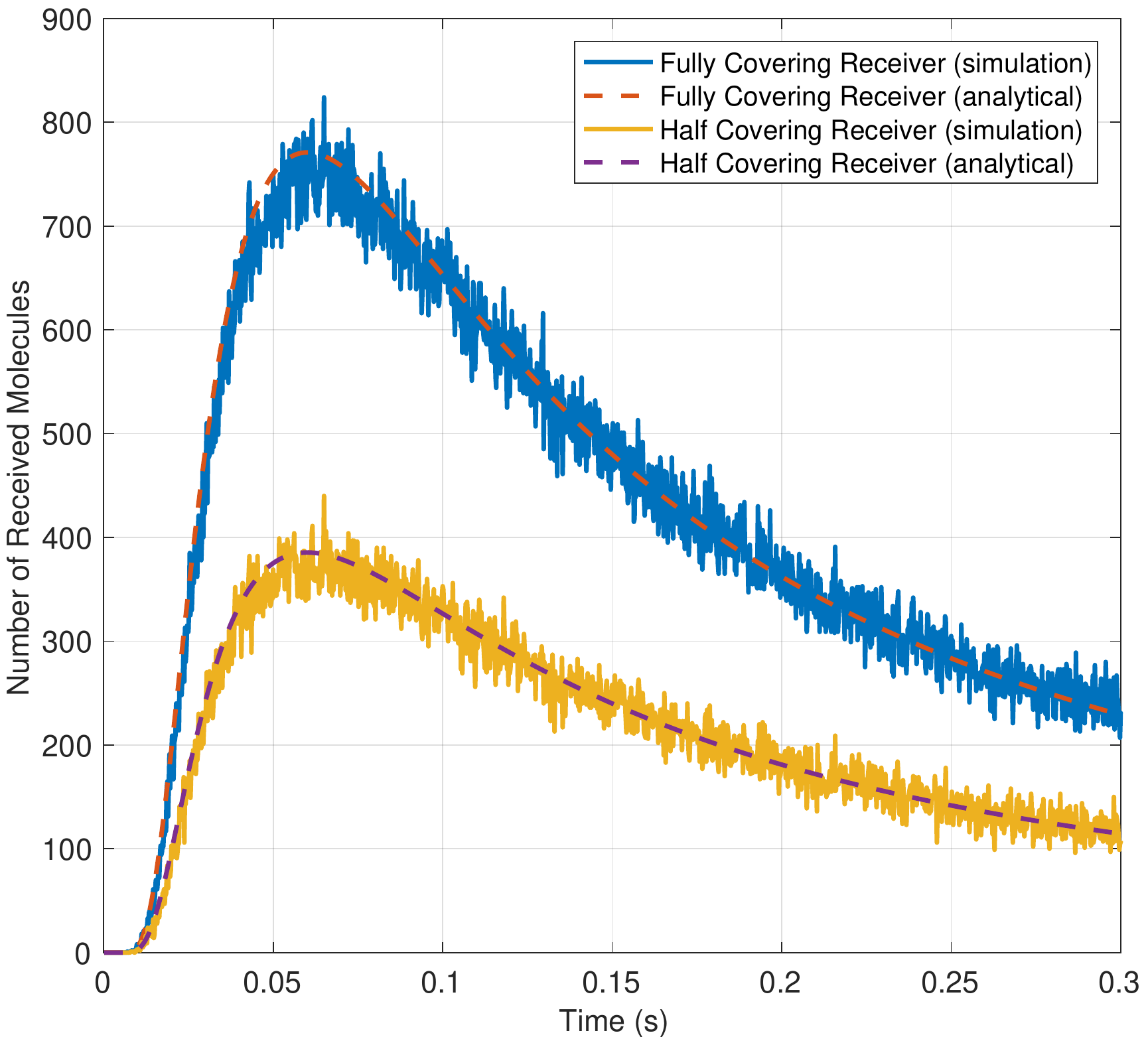}
        \label{fig_patch_r3_d6_D100}}        
	\end{center}
   \caption{Time versus number of received molecules for both fully and partially covering receivers} 
   \label{fig_patch_receivedMolecules}
\end{figure*}
%%%%%%%%%%%%%%%%%%%%%%%%%%%%%%%%%%%%%%%%%%%%%%%%%%%%

%\lipsum[1-5]
In the literature, the receiver cells are mostly considered to fully cover the vessel in two dimensions (excluding the dimension that MMs propagate) in vessel-like environments~\cite{TURAN2018_MOLEye}. Therefore, the communication channel reduces to 1D channel. However, we believe that designing the receiver cell as partially covering rather than designing it as fully covering has two important advantages: preventing potential health problems and realizing relay nodes in the conventional communication systems.

While building a nanomachine for blood vessels, the risk of vascular occlusion should be considered and minimized in the deployment process independent from the application. Even a small inattentiveness may cause serious health problems. Vascular occlusion is a common and extremely important cause of clinical illness which is the reason of serious amount of deaths in the world~\cite{Kumar2017_Pathology}. To this end, partially covering receiver should be used to avoid causing vascular occlusion and consequently serious health problems.

%%relay'den bahset
The other reason for considering the partially covering receiver is to use the relay node concept in MCvD. Since the maximum amount of distance to propagate through the vessel in MCvD is limited, relay nodes can be used to amplify the signal. By doing so, MCvD can be used in order to communicate with much higher distances, which is not possible without using relay nodes. It is also biologically friendly since the receiver cell (also the transmitter cell for the next link) has a partially covering structure.

Due to both extending the communication range and preventing the health problems, considering and modeling the partially covering receiver is at a great importance. Therefore, we propose an approximation for the channel model of MCvD in vessel-like environments by considering the scenarios where the received molecules are dispersed homogeneously on the cross-section of the vessel.

As shown in Table~\ref{tbl_ks_tests}, when we have $d/r_{\text{v}}$ greater than a specific value, the hitting molecules are distributed uniformly. Therefore, in our proposed channel model, we scale the 1D formulation by the ratio of the partially covering receiver area to the whole cross-sectional area as follows:
%%%%%%%%%%%%%%%%%%%%%%%%%%%%%%%%%%%%%%%%%%%%%%%%%%%%
\begin{align}
\begin{split}
F_{\text{hit}}(t) &= N^{\text{Tx}} \,\, \Phi(\Omega) \,\, \text{erfc}(\frac{d}{\sqrt{4Dt}})\\
\Phi(\Omega) &= \frac{\text{A}(\Omega)}{\pi r_{\text{v}}^{2}}
\end{split}
\label{eq_1DFormula}
\end{align}
%%%%%%%%%%%%%%%%%%%%%%%%%%%%%%%%%%%%%%%%%%%%%%%%%%%%
where $\text{A}(\Omega)$ represents the area of the partially covering receiver $\Omega$. In order to validate~(\ref{eq_1DFormula}), three different environmental conditions (that passed K-S test) are considered. As can be seen in Fig.~\ref{fig_patch_receivedMolecules}, $t_p$ (which is equal to $\frac{d^2}{6D}$) increases with the increasing $d$ or decreasing $D$ and vice versa. Also, simulation results of both fully and partially receivers perfectly follow the formula stated in~(\ref{eq_1DFormula}), which is the channel modeling goal of this study.

%%%%%%%%%%%%%%%%%%%%%%%%%%%%%%%%%%%%%%%%%%%%%%%%%%%%
%%%%%%%%%%%%%%%%%%%%%%%%%%%%%%%%%%%%%%%%%%%%%%%%%%%%
%%%%%%%%%%%%%%%%%%%%%%%%%%%%%%%%%%%%%%%%%%%%%%%%%%%%
%%%%%%%%%%%%%%%%%%%%%%%%%%%%%%%%%%%%%%%%%%%%%%%%%%%%

%%%%%%%%%%%%%%%%%%%%%%%%%%%%%%%%%%%%%%%%%%%%%%%%%%%%
%%%%%%%%%%%%%%%%%%%%%%%%%%%%%%%%%%%%%%%%%%%%%%%%%%%%
%%%%%%%%%%%%%%%%%%%%%%%%%%%%%%%%%%%%%%%%%%%%%%%%%%%%
%%%%%%%%%%%%%%%%%%%%%%%%%%%%%%%%%%%%%%%%%%%%%%%%%%%%
\section{Conclusion and Future Work}
%\lipsum[1-2]
In this paper, we have analyzed hitting location distribution of MMs in vessel-like environments. Also, potential application and importance of partially covering receiver is emphasized. Moreover, an approximation for the channel model is given for the cases with sufficiently high $d/r_{\text{v}}$ values, i.e., uniformly distributed molecule locations.

Two dimensional distribution of hitting molecules are analyzed in two parts, namely angular and radial distributions. While analyzing distribution of MM in radius, radial distribution of MMs are analyzed using K-S test for two different significance level, namely 1\% and 5\%. In order to increase the reliability of the results, More than 100 different environmental conditions are analyzed. All in all, it is shown that molecules that have tendency to spread uniformly in the receiver area beyond certain ratio of $d/r_{\text{v}}$. Using these results, we proposed a channel model for partially covering receiver in vessel-like environments and we verified the channel model by simulations.

As a future work, we plan to find the formula that gives the exact distribution of the hitting molecules that eliminates the necessity of having a constraint on $d/r_{\text{v}}$.

%which may create potential application area for patch receiver which can be used as relay nodes and avoid causing vascular occlusion.

% conference papers do not normally have an appendix

% use section* for acknowledgment
%\section*{Acknowledgment}
%This research was partially supported by the Scientific and Technical Research Council of Turkey (TUBITAK) under Grant number 116E916 and by Generalitat de Catalunya of the Secretariat for Universities and Research of the Ministry of Business and Knowledge of the Government of Catalonia via the Beatriu de Pinos program.

% trigger a \newpage just before the given reference
% number - used to balance the columns on the last page
% adjust value as needed - may need to be readjusted if
% the document is modified later
%\IEEEtriggeratref{8}
% The "triggered" command can be changed if desired:
%\IEEEtriggercmd{\enlargethispage{-5in}}

% references section

\bibliographystyle{IEEEtran}
\bibliography{Uniform_2018}

\end{document}